\documentclass[preprint]{aastex63}
\accepted{July 10, 2021}
\submitjournal{ApJL}

\shorttitle{The solar minimum 24/25}
\shortauthors{Li et al.}
\begin{document}

\title{Is Solar Minimum 24/25 Another Unusual One?}

\correspondingauthor{Xueshang Feng}
\email{fengx@spaceweather.ac.cn}

\author[0000-0002-1732-0196]{Huichao Li}
\affiliation{Institute of Space Science and Applied Technology, Harbin Institute of Technology Shenzhen, Shenzhen 518055, China}
\affiliation{SIGMA Weather Group, State Key Laboratory of Space Weather, National Space Science Center, Chinese Academy of Sciences, Beijing 100190, China}

\author[0000-0001-8605-2159]{Xueshang Feng}
\affiliation{Institute of Space Science and Applied Technology, Harbin Institute of Technology Shenzhen, Shenzhen 518055, China}
\affiliation{SIGMA Weather Group, State Key Laboratory of Space Weather, National Space Science Center, Chinese Academy of Sciences, Beijing 100190, China}

\author{Fengsi Wei}
\affiliation{Institute of Space Science and Applied Technology, Harbin Institute of Technology Shenzhen, Shenzhen 518055, China}
\affiliation{SIGMA Weather Group, State Key Laboratory of Space Weather, National Space Science Center, Chinese Academy of Sciences, Beijing 100190, China}



\begin{abstract}
{
The solar minimum 23/24 is considered to be unusual because it exhibits features that differ notably from those commonly seen in pervious minima. In this letter, we analyze the solar polar magnetic field, the potential-field solution of the solar corona, and the in-situ solar wind measurements to see whether the recent solar minimum 24/25 is another unusual one. While the dipolar configuration that are commonly seen during minimum 22/23 and earlier minima persist for about half a year after the absolute minimum of solar cycle 24, the corona has a morphology more complex than a simple dipole before the absolute minimum. The fast solar wind streams are less dominant than minimum 23/24. The IMF strength, density and mass flux that are historically low in the minimum 23/24 are regained during minimum 24/25, but still do not reach the minimum 22/23 level. From the analysis of this Letter, it seems that the minimum 24/25 is only partially unusual, and the recovery of the commonly minimum features may result from the enhancement of the polar field.}
\end{abstract}

\keywords{Solar Cycle; Solar Corona; Solar Wind; Interplanetary Magnetic Fields}

\section{Introduction}

{In solar minimum 22/23 and earlier ones,} the picture of the corona and heliosphere during the solar minimum is characterized by a dipolar configuration with limited tilt with respect to the solar rotational axis. The heliospheric current sheet (HCS) lies near the solar equator with a flat shape. The interplanetary space is dominated by open magnetic flux and solar wind emitted from the large polar coronal holes. While the high-latitude region is occupied by the uniform high-speed solar wind, slow and variable solar wind is seen near the ecliptic plane \citep{Gazis1996,Gibson2001,Richardson2008}. 

However, the solar minimum 23/24 exhibits features that differ notably from {features commonly seen in minimum 22/23 and earlier ones}. While the sunspot activity is historically low, the large-scale corona morphology is more complex than a simple dipole. The polar coronal holes are less dominant than those in previous solar minima, but the low-latitude coronal holes, which are not commonly seen during the previous solar minimum, are relatively large and persist many rotations \citep{Abramenko2010,Hewins2020}. As a result, high-speed solar wind streams originated from these low-latitude coronal holes occur more frequently in the near-ecliptic region, generating periodic forcing of the heliosphere and Earth \citep{Gibson2009,Ram2010,Toma2011}. Nevertheless, in situ measurements obtained near the ecliptic \citep{Jian2011} and in high latitude \citep{McComas2013} both indicate that the Sun produces weaker output, as the solar wind mass flux and the interplanetary magnetic field (IMF) strength are lower than those during minimum 22/23. Meanwhile, the HCS stays elevated when the sunspot number (SSN) is already close to its ultimate minimum \citep{Toma2010ASPC,Zhao2013}. Because of these distinctive features, the solar minimum 23/24 is considered as an unusual minimum and attracts considerable research efforts (e.g., see \citet{Gibson2011}, \citet{Cranmer2010} and \citet{Luhmann2012} and references therein). These works achieve the consensus that the peculiar behaviors of the corona and heliosphere are attributed to the weak polar field of the Sun \citep{Wang2009,Luhmann2009}.

\begin{figure}
        \centering
        \centerline{\includegraphics[width=\textwidth]{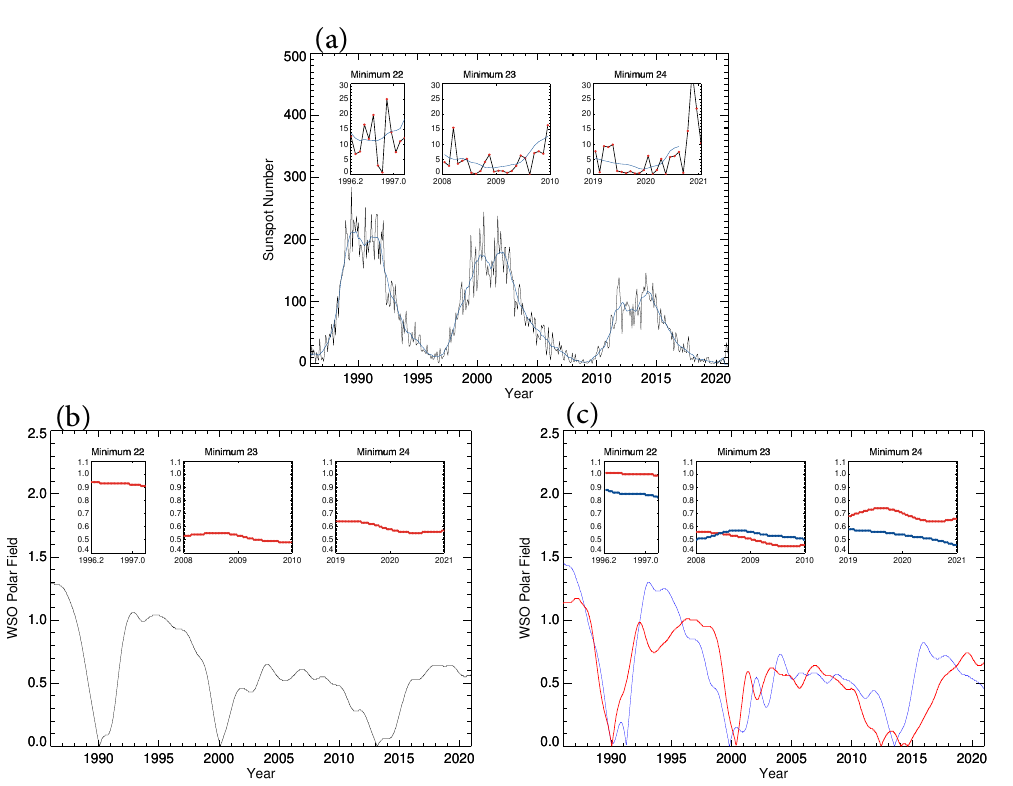}}
        \caption{(a) Evolution of the sunspot number (SSN). The monthly and the 6-month smoothed SSN are represented by black and blue lines. (b) Evolution of the WSO average polar field strength. (c) Evolution of the WSO polar field strength in the northern (red) and southern (blue) poles.}
        \label{fig:sunspot_polar}
\end{figure}

Following this unusual minimum, the Sun enters solar cycle 24, one of the weakest solar cycles since solar cycle 14 \citep{Bisoi2020}. As can be seen from the SSN shown in Figure \ref{fig:sunspot_polar} (a), the ``mini'' maximum of solar cycle 24 is significantly less vigorous than the previous two maxima. Since the solar cycle 24 has this special character, we ask the question that {if the solar minimum 24/25 is another unusual one}. In this Letter, we try to, at least partially, answer this question by comparing minimum 24/25 with minimum 22/23 that {has the common minimum features}, and minimum 23/24 that appears to be unusual.

\section{Data and Method}

In order to analyze the character of the solar minimum, it is a common approach to choose a time period around absolute minimum when the solar activity is continuously low. {Note that the absolute minimum refers to the month when the 13-month smoothed monthly SSN reaches its minimum \footnote{http://www.sidc.be/silso/cyclesminmax}}. For the three minima compared in this study, the absolute SSN minima are passed on September 1996, December 2008 and December 2019, respectively. A period of one year from March 1996 to February 1997 is chosen for minimum 22/23. Since minimum 23/24 is long and deep, a period of two years or more is usually used to illustrate the character of this minimum in full context \citep[e.g.][]{Toma2011,Gibson2011}. Therefore, we choose a two-year period, 2008 and 2009, for minimum 23/24 in this study. As can be seen from Figure \ref{fig:sunspot_polar}a, the sunspot active level is also low in the two-year period around the absolute minimum of solar cycle 24, and we therefore also choose a two-year period from 2019 to 2020 for minimum 24/25. The SSN data is obtained from Sunspot Index and Long-term Solar Observations (SILSO).

The potential field source surface (PFSS) model has been shown to have the ability to describe the realistic large scale coronal fields with reasonable accuracy \citep[e.g.,][]{Neugebauer1998,Schrijver2003,Riley2006,Wang2009,Badman2020}.  The PFSS code used in this study was developed by \citet{Jiang2012}\footnote{Available at \url{https://sourceforge.net/projects/glfff-solver/}}, and the source surface is set at the commonly used 2.5 $\rm R_s$ ($\rm R_s$, solar radius). The positions of the coronal holes are found by tracing the magnetic field lines from solar surface to the source surface. The position of the HCS at the source surface are determined by the position where the radial field ($\rm B_r$) change its sign. The warp of the HCS is measured by the integrated slope \citep{Zhao2013}:

\begin{equation}
\label{equ:SL}
SL=\frac{1}{N}\sum {\Big|} \frac{\partial \theta_i}{\partial \phi_i} {\Big|}
\end{equation}
where $\theta_i$ and $\phi_i$ are the latitude and longitude of the grid that the HCS lies on, and the derivative is calculate by the difference between neighbour points. A higher SL means a more warped HCS. As input to the PFSS model, we use the Wilcox Solar Observatory (WSO) synoptic maps for minimum 22/23, and the Global Oscillation Network Group (GONG) synoptic maps for minima 23/24 and 24/25. The GONG maps available since 2006 have higher spatial resolution and less data gaps. The synoptic maps are generated for each Carrington rotation (CR). The polar fields are analyzed by using WSO polar field observations \footnote{\url{http://wso.stanford.edu/Polar.html}}, which provide the most untouched measure of the solar polar field strength. 

The near-Earth insitu data is obtained from the OMNI data base.  We use the Morlet wavelet to analyze the periodic behavior of the near-Earth solar wind \citep{Torrence1998}. We employ the relative occurrence difference ratio of the IMF polarity defined by $\rm R_{TA}=(T-A)/(T+A)$ to see the displacement of the HCS from the heliographic equator \citep{Mursula2011,Koskela2018}. The value of T (A) are calculated by the total number of OMNI hourly IMF data of $B_y-B_x<0$ ($B_y-B_x>0$) in the GSE coordinate system. The annual $\rm R_{TA}$ is calculated to remove the geometric effect like the Russell–McPherron effect. During minima 22/23 and 24/25, the northern (southern) heliosphere is dominated by A (T) polarity, therefore a positive (negative) $\rm R_{TA}$ means the northward displacement of the HCS. Inversely, a positive (negative) $\rm R_{TA}$ means the southward (northward) displacement of the HCS in minimum 23/24, due to the reverse of polar polarity.

\section{Result and Discussion}

In Figure \ref{fig:sunspot_polar}, we present the monthly sunspot number (SSN) and the polar field strength of the two poles (Figure \ref{fig:sunspot_polar} (c)) and their average (Figure \ref{fig:sunspot_polar} (b)). A zoom-in view is provided for the three minima. While minimum 23/24 is characterized by the low activity level, the Sun is also relatively quiet during minimum 24/25, showing lower activity level than minimum 22/23, except for the end of 2020 when the SSN raises sharply. Despite the low activity level during the solar cycle 24, the polar field strength of minimum 24/25 is 13\% stronger than that during minimum 23/24 in average, but still 34\% weaker than minimum 22/23.  The north-south asymmetry of the polar field is more evident during minimum 24/25 than during minimum 23/24, with the northern polar field (red line) about 30\% stronger than the south (blue line). A similar north-south asymmetry is found during minimum 22/23.  It has been noticed that in solar cycle 24 the northern polar field reverses earlier than the southern polar field, but reverses for multiple times \citep{Sun2015,Janardhan2018}. As can be seen from Figure \ref{fig:sunspot_polar}c, the northern polar field is actually weaker than the southern polar field during most of the solar cycle 24. After the polar reversal around 2003-2004, the southern polar field builds up more quickly but starts to decrease earlier than the northern polar field, resulting in weaker strength during the minimum. 

\begin{figure}
        \centering
        \centerline{\includegraphics[width=\textwidth]{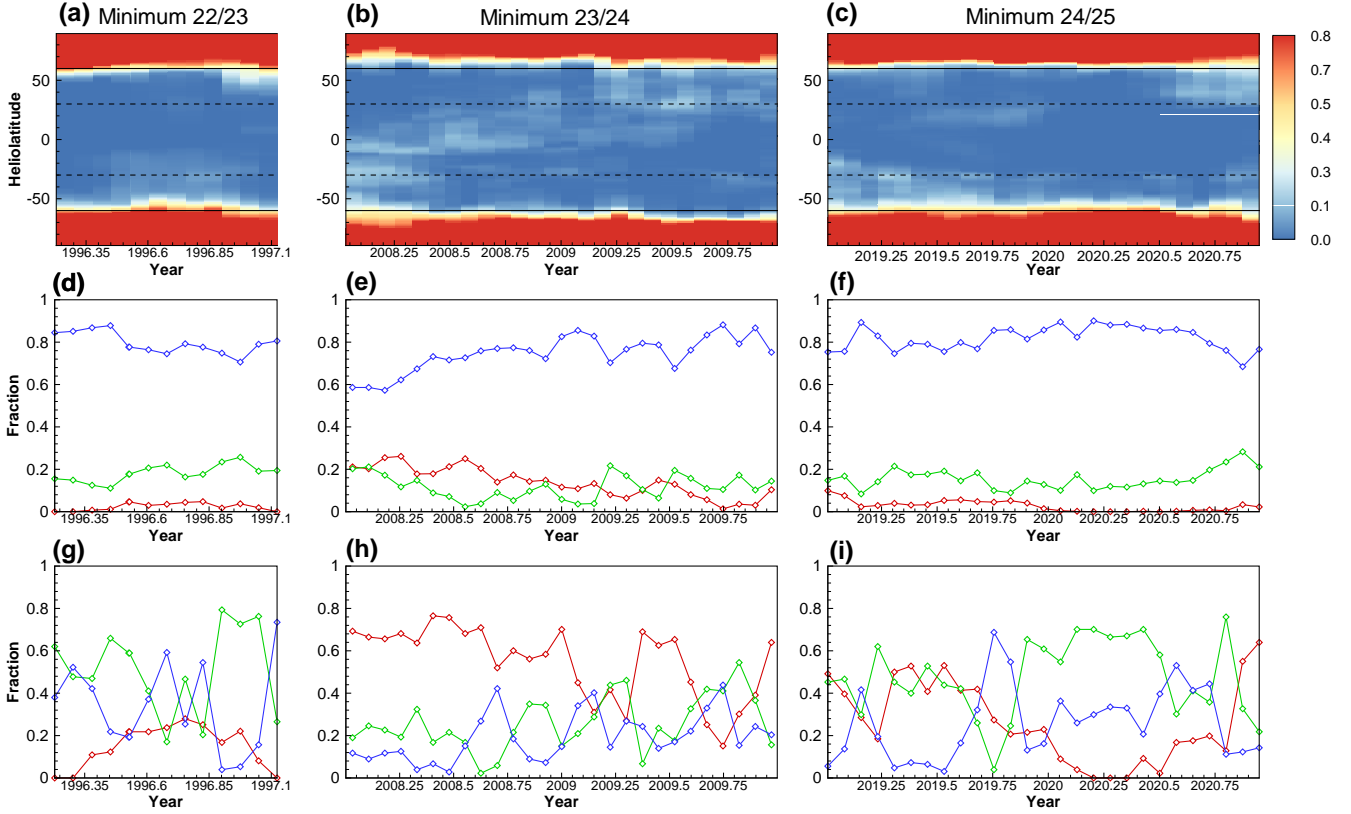}}
        \caption{(a)-(c) The latitudinal distributions of the PFSS-derived coronal hole integrated over longitude, represented by the fraction
of longitude grids that are identified as coronal hole in each latitude grid. The solid and dashed lines separate high/mid and mid/low latitude respectively. (d)-(f) The fraction of coronal hole distributed in low (red), mid (green) and high (blue) latitude. (g)-(i) The fraction of the foot points of ecliptic magnetic field lines that are at the low (red), mid (green) and blue (red) latitude.}
        \label{fig:CH}
\end{figure}

In Figure \ref{fig:CH}(a)-(c), we present the latitudinal distributions of the PFSS-derived coronal hole integrated over longitude. We divide the solar surface into the high-latitude region above $\pm 60^\circ$, the low-latitude region between $\pm 30^\circ$ and the mid-latitude region between them. The fraction of the coronal hole distributed in each region is plotted in Figure \ref{fig:CH}(d)-(f). The low-, mid- and high-latitude regions are represented by blue, green and red lines, respectively. We also trace back along the near-ecliptic open magnetic field lines from the source surface to the solar surface, and plot the fraction of foot points in each region in Figure \ref{fig:CH} (g)-(i) to illustrate the source position of the solar wind that sweeps the Earth. The minimum 22/23 sees the {nearly dipolar} distribution of the coronal holes. The solar surface is dominated by the polar coronal hole in high latitude and their extension in the mid latitude. The majority of the near-ecliptic field lines are traced back to the mid- and high-latitude. In contrast, during minimum 23/24 the band of isolated low-latitude coronal hole persisted until 2009.5, taking a considerable fraction of the coronal hole area. These low-latitude coronal holes are the major source of the near-ecliptic solar wind, since most near-ecliptic field lines originate from the low-latitude region. For minimum 24/25, the low-latitude coronal holes are less predominant than minimum 23/24. The fraction of the low-latitude coronal hole area in Figure \ref{fig:CH}(f) is at the same level as minimum 22/23. However, an isolated band of low-latitude coronal hole is found in the second half of 2019 in Figure \ref{fig:CH} (c), resulting in about half of the near-ecliptic field lines originating from low-latitude. The mid-latitude coronal holes have a   north-south asymmetric distribution during 2019, and are heavily distributed in the south. After the absolute minimum at the end of 2019, the corona exhibit the classic dipolar configuration that resembles minimum 22/23, as the low-latitude coronal holes nearly disappear and most of the near-ecliptic field lines originate from the mid- and high-latitude region. The corona is dominated by the polar coronal hole and its low-latitude extension until the end of 2020, when the solar activity of the new cycle raises and brings in mid-latitude coronal holes. 

\begin{figure}
        \centering
        \centerline{\includegraphics[width=\textwidth]{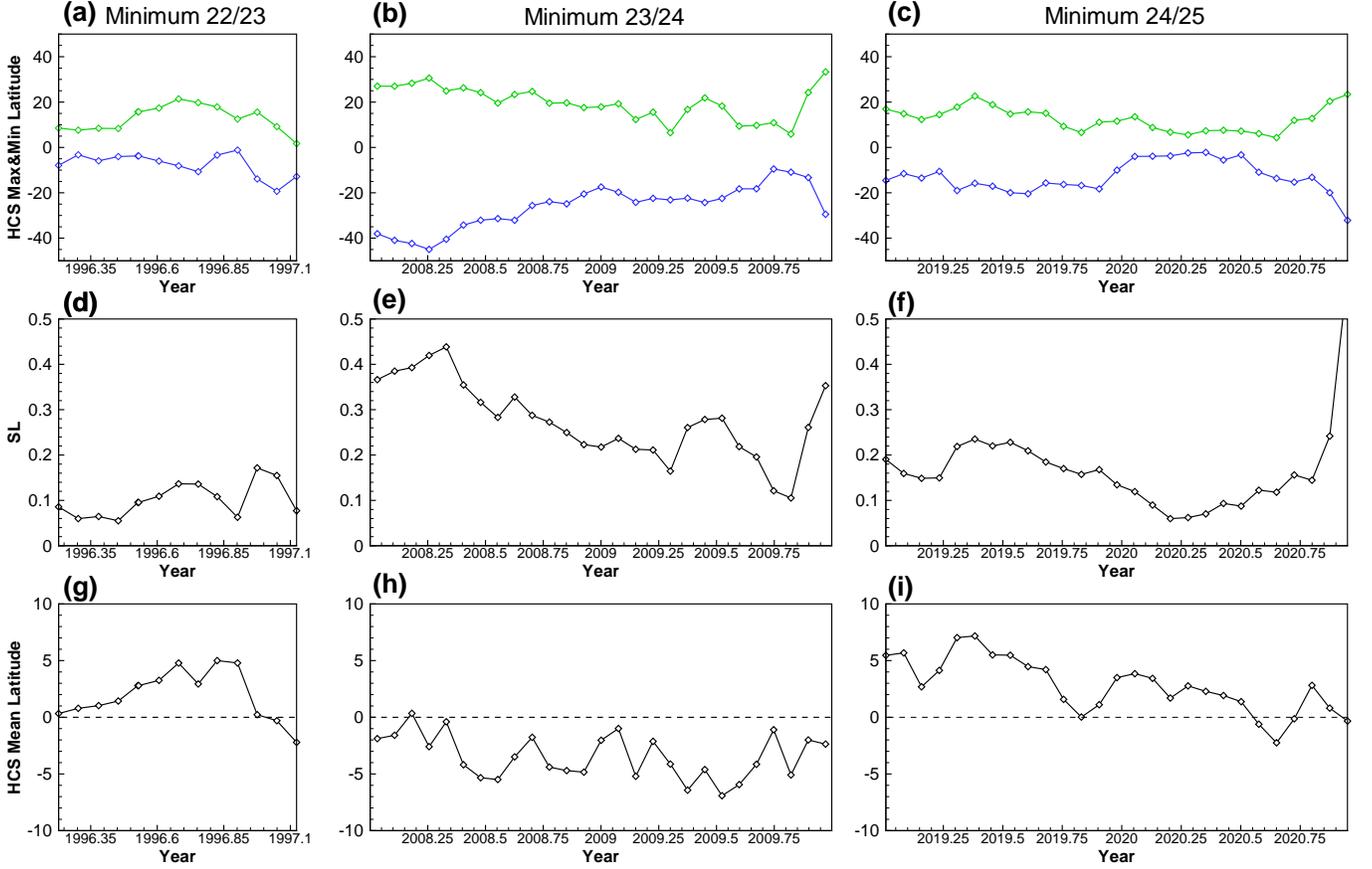}}
        \caption{(a)-(c) The maximum northern (green) and southern (blue) latitudinal extent of the HCS. (d)-(f) The integrated slope (SL) of the HCS, calculated by Equation (\ref{equ:SL}). (g)-(i) Longitudinal average latitude of the HCS. The dashed line marks the position of zero value. }
        \label{fig:HCS}
\end{figure}

Figure \ref{fig:HCS} (a)-(c) show the maximum northern (green) and southern (blue) latitudinal extent of the HCS . Figure \ref{fig:HCS} (d)-(f) present the SL of the HCS calculated by Equation (\ref{equ:SL}). During minimum 22/23, the HCS is confined approximately within latitude $\pm 20^\circ$, with SL varying around 0.1 and never exceeding 0.2. During most of the minimum 23, the latitudinal extension of the HCS is larger than that of minimum 22/23. The maximum is seen around year 2008.25, being about $-40^\circ$ in the south and $30^\circ$ in the north. The SL is also much larger than  that of minimum 22 in general, indicating a very warped HCS that extends to relatively high latitude. For minimum 24/25, the latitudinal extension of the HCS is significantly smaller than the counterpart of minimum 23/24. Nevertheless, except for the first half of year 2020, the HCS still extends to latitude slightly higher than that of minimum 22/23, and is more warped than minimum 22/23. During the first half of 2020, small HCS latitudinal  extension and small SL value are seen at the same time, indicating that a flat HCS lies close to the equator, which resembles the classic dipolar morphology of the HCS during solar minimum. In the end of 2020, as the sunspot activity of the new cycle raises rapidly, the dipolar morphology of the HCS vanishes and a more complex HCS structure occurs. 

\begin{figure}
        \centering
        \centerline{\includegraphics[width=\textwidth]{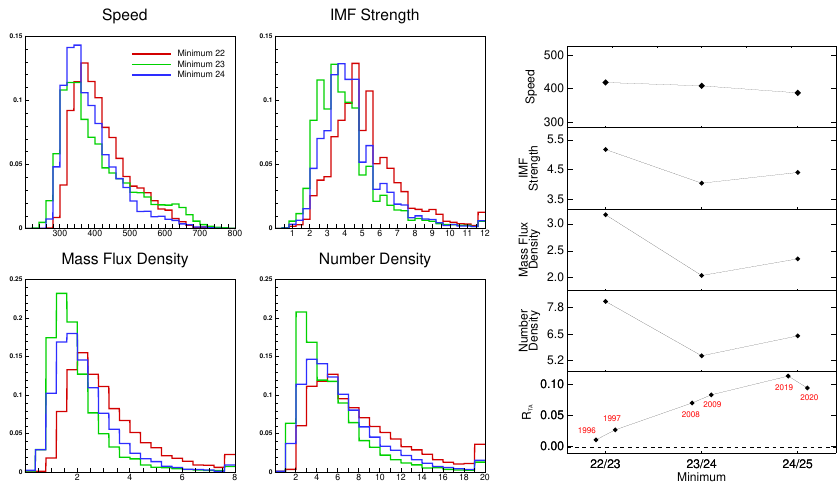}}
        \caption{Distributions (left) and averages (right) of OMNI solar wind speed (km/s), IMF strength (nT), mass flux density ($\rm 10^3\ cm^{-3}km/s$) and number density ($\rm cm^{-3}$). The change of annual $\rm R_{TA}$ ratios is also shown in the right panel. The bin sizes of the distributions are 20 km/s for speed, 0.5 nT for IMF strength, $\rm 0.4\times 10^3\ cm^{-3}km/s$ for mass flux density, and 1 $\rm  cm^{-3}$ for number density.}
        \label{fig:distribution}
\end{figure}

From the beginning of the space age to the minimum 23/24, the HCS has been found to have a general trend of shifting southward from the heliographic equator during solar minimum from both in situ measurements and global modeling \citep{Mursula2003,Zhao2005,Koskela2018}. Although the southward shift of the HCS is observed between 1994 and 1995 \citep{Erdos2010}, this trend is interrupted in the study interval of minimum 22/23 in this paper. The longitudinal average of the HCS latitude is dominated by positive value, and the annual  $\rm R_{TA}$ ratios for 1996 and 1997 are {both positive as can be seen in Figure \ref{fig:distribution}}, indicating that the HCS shifts northward. The southward shift of HCS recurs during minimum 23/24, as the average latitude of HCS is dominated by negative value, and the annual $\rm R_{TA}$ ratios for 2007 and 2008 are both positive. A prediction is made at the beginning of solar cycle 24, inferring that the HCS would shift northward in minimum 24/25, which ends the tendency of the southward shift of the HCS \citep{Mursula2011}. From Figure \ref{fig:HCS} (i), we do see an HCS that remains shifting northward during 2019 and 2020.  The annual $\rm R_{TA}$ ratios are {both positive} for 2019 and 2020. The tendency of the northward shift is thus reflected by both coronal modelling results and near-Earth in situ data.

{The change of synoptic map sources may lead to some differences in the PFSS results. We also calculate the results with the filled version of WSO maps, in which the missing data are filled by interpolation from adjacent CRs. Figures from the WSO results showing the same information as that of Figures \ref{fig:CH} and \ref{fig:HCS} are presented in the support material. It can be seen that although the WSO results tend to give larger fraction of coronal hole area in the mid-latitude, the main findings from the GONG results, including the evolution of coronal hole distributions, the dominance of low/mid/high latitude coronal holes in footpoints of the near-ecliptic field lines, and the evolution of the HCS structures, are still valid.} 

%The north-south asymmetry induced by the shift of the HCS is stronger in 2019 than in 2020, as both the average latitude of HCS and the T-A/T+A are larger.

\begin{figure}
        \centering
        \centerline{\includegraphics[width=\textwidth]{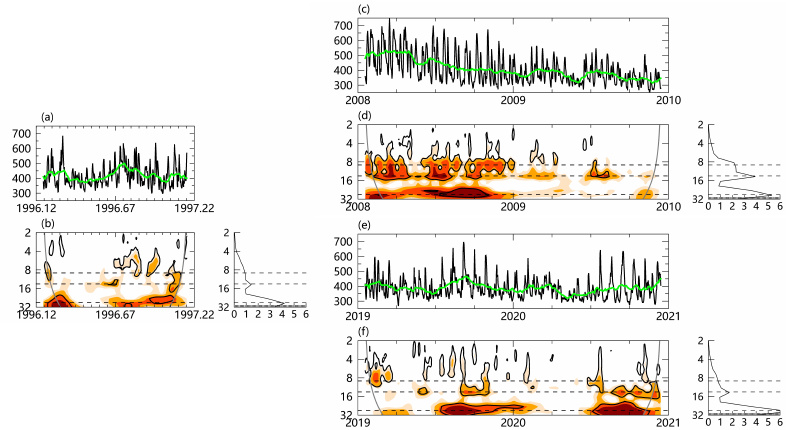}}
        \caption{(a) The evolution of the daily solar wind speed (black) and its 27-day averaged value (green) during minimum 22/23. (b) The wavelet power spectrum (left) and the global wavelet spectrum (right) of the daily solar wind speed for minimum 22/23. The dashed lines mark the periodicities at 27, 13.5 and 9 days. The 95\% significant level is marked by the solid black contour lines. (c)-(d) and (e)-(f) provide the same information as (a)-(b) but for minima 23/24 and 24/25, respectively.}
        \label{fig:wavelet}
\end{figure}

Figure \ref{fig:wavelet} presents the time series of the daily mean solar wind speed and the wavelet power spectrum of the solar wind speed. During minimum 22/23, the solar wind exhibits periodicity of solar rotation (27 days) at the very beginning and the second half of the study interval. During minimum 23/24, relatively structured fast solar wind streams persist through 2008, with most of their maximum speed above 600 km/s. Periodicities at both solar rotation (27 days) and its subharmonic periods (13.5, 9 days) are observed for continuously long patches with high spectral intensities. The periodicity at solar rotation indicates that the fast streams recur in each rotation, implying that the low-latitude coronal holes generating these streams persist for many rotations in this interval. The subharmonic periods implying two or three major low-latitude coronal hole regions are longitudinally separated, resulting in a pattern of two-three peaks in the speed profiles \citep{Luhmann2009,Ram2010,Li2018}. In 2009, the solar wind speed starts to decay and the strong periodicities disappear. This change may be explained by what can be observed from Figure \ref{fig:CH} (b) and (e). The distributions of the low- and mid-latitude coronal holes are fickle in 2009, and the low-latitude coronal holes gradually diminish. For the recent solar minimum 24/25, strong fast solar wind streams with maximum speed higher than 600 km/s are observed between yr 2019.5 and 2019.9, when the band of isolated low-latitude coronal hole in Figure \ref{fig:CH} (c) is seen. Significant periodicity at solar rotation is continuously observable during the occurrence of the fast streams, being similar to what has been seen in 2008. However, the subharmonic solar rotational periodicity is not evident, implying there may be only one major fast stream in each rotation. Between 2020.0 and 2020.5, as the solar corona evolves into the dipolar state, the slow solar wind becomes dominant, and no periodicity pattern can be seen in the wavelet spectrum. A similar situation can be found in minimum 22/23 around the mid of 1996. In the second half of 2020, the fast streams are observed again, with apparent one-rotation periodicity and observable half-rotation periodicity, {indicating that a two-peak structure recurs in each rotation.}

Figure \ref{fig:distribution} compares the distributions {and averages} of solar wind parameters in the three minima. The shapes of the distributions of speed for minima 22 and 24 look similar, but the occurrence of high-speed solar wind above 500 km/s in minimum 24/25 is the lowest among the three minima. The distribution of minimum 23/24 is characterized by the  high-speed tail above 600 km/s, but such high-speed tail is absent in minima 22/23 and 24/25. As can be seen in Figure \ref{fig:wavelet} (a), (c) and (e), the 27-day average speed represented by the green line fluctuates around 400 km/s in minima 22/23 and 24/25. In contrast, it decreases gradually from about 500 km/s to about 350 km/s in minimum 23/24, following the decaying of the low-latitude coronal hole seen in Figure \ref{fig:CH}(e). The {average solar wind speed of minimum 24/25 is lower} than that of the two previous minima, and even lower than that of minima 20/21 and 21/22 \citep{Jian2011}.  The average IMF strength during minimum 24/25 is weaker than that of minimum 22/23, but stronger than that of minimum 23/24. {The strengthen of IMF from minimum 23/24 to 24/25 ends the trend of IMF weakening from minimum 21/22 to 23/24 \citep{Jian2011}}, and may also be in connection with the increase of the solar polar field.  The distribution of mass flux density and number density of minimum 24/25 have occurrence of large value higher than that of minimum 23/24, and the occurrence of small value higher than that of minimum 22/23. {The average number density and mass flux density of minimum 24/25 lie between those of minima 22/23 and 23/24.}

\section{Concluding Remarks}

{

In this Letter, we analyze the features of coronal structure and solar wind parameters during minimum 24/25 by comparing them with those of the previous two minima. In the perspective of corona, while the dipolar configuration that are commonly seen during minimum 22/23 and earlier minima persist for about half a year, the corona exhibits a morphology more complex than a simple dipole before the absolute minimum, showing unusual features seen in minimum 23/24. During that period, the low-latitude corona hole remains an important source of the near-ecliptic solar wind, and the HCS is more warped and extended to higher latitude than that of minimum 22/23. Meanwhile, the north-south asymmetry of the corona during minimum 24/25 is reflected by the distribution of coronal holes and the northward displacement of the HCS. In the perspective of near-Earth solar wind, in minimum 24/25, the IMF strength, density and the mass flux that are historically low in minimum 23/24 are regained, but still do not attain their minimum 22/23 level. The fast streams are less dominant than in minimum 23/24, and the distribution of speed more resemble the common solar minimum profile. The average speed is the lowest among the historic record of solar wind during minimum. However, clear solar-rotation periodicity can be found in the speed when the low-latitude coronal hole occurs. 

From the analysis of this Letter, it seems that the minimum 24/25 is only partially unusual. The recovery of the common minimum features may result from the enhancement of the polar field. While this Letter provide an overview of the corona structure and the solar wind parameter during minimum 24/25, a more detailed understanding of minimum 24/25 may come from coordinated observation and modeling efforts focusing on some particular periods, which resembles the successful Whole Heliosphere Interval (WHI) of minimum 23/24 \citep{Gibson2011}. Meanwhile, it is also interesting to look into how the change of the corona and solar wind impacts the terrestrial environment, like the periodic forcing of the ionosphere and thermosphere. 

Since the polar field of the Sun plays a fundamental role in determining the magnetic configuration of the coronal and the property of the solar wind during solar minimum, it seems that the polar field strength can be a concise indicator of each minimum’s unusualness. However, it is difficult to properly set a threshold to judge whether or not a certain minimum is unusual. The “unusual” impression is based on the observation and modelling efforts only for only a few minima. Maybe each minimum has its own features, and needs to be characterized by comprehensive study of long time dataset.

}

\acknowledgments

The work is jointly supported by the National Natural Science Foundation of China (grants  41874202, 42030204, 41861164026, 41731067) and the Guangdong Basic and Applied Basic Research Foundation (2019A1515011067). The present work is also partially supported by the National Key Scientific and Technological Infrastructure project ``Earth System Science Numerical Simulator Facility'' (EarthLab). This work utilizes GONG data from NSO, which is operated by AURA under a cooperative agreement with NSF and with additional financial support from NOAA, NASA, and USAF. Wilcox Solar Observatory data used in this study was obtained via the web site \url{http://wso.stanford.edu}. We acknowledge use of NASA/GSFC's Space Physics Data Facility's OMNIWeb and CDAWeb service for providing the observed {\it in-situ} data used in this paper. Sunspot number data is obtained from  WDC-SILSO, Royal Observatory of Belgium. Wavelet software was provided by C. Torrence and G. Compo, and is available at URL: \url{http://atoc.colorado.edu/research/wavelets/}.

\bibliography{minimum_compare}
\bibliographystyle{aasjournal}

\end{document}